\documentclass[1p,number]{elsarticle} 
\usepackage{graphicx,graphics}
\usepackage{citesort}
\usepackage{rotating,rotate}
\usepackage{epsfig}
\usepackage{exscale}
\usepackage{natbib}
\usepackage{amsmath,amssymb}
\usepackage{multirow}
\usepackage{latexsym}
\newcommand{\bc}{\begin{center}}
\newcommand{\ec}{\end{center}}
\newcommand{\bfi}{\begin{figure}}
\newcommand{\efi}{\end{figure}}
\newcommand{\bm}{\begin{minipage}}
\renewcommand{\em}{\end{minipage}}
\newcommand{\be}{\begin{equation}}
\newcommand{\en}{\end{equation}}
\newcommand{\bea}{\begin{eqnarray}}
\newcommand{\eea}{\end{eqnarray}}
\newcommand{\bi}{\begin{itemize}}
\newcommand{\ei}{\end{itemize}}
\newcommand{\no}{\nonumber}

\newcommand{\ov}[1]{\overline{#1}}
\newcommand{\bbr}{{\it B{\footnotesize A}B{\footnotesize AR}}}
\def\b         {\ensuremath{\mathcal{B}}}
\def\bb        {\ensuremath{\mathcal{B}\overline{\mathcal{B}}}}
\def\eepp        {\ensuremath{e^+ e^- \!\rightarrow p\overline{p}\,}}

\def\eebb        {\ensuremath{e^+ e^- \!\rightarrow \mathcal{B}\mathcal{\overline{B}}\,}}
\def\eell        {\ensuremath{e^+ e^- \!\rightarrow \Lambda\overline{\Lambda}}}
\def\eelclc      {\ensuremath{e^+ e^- \!\rightarrow \Lambda_c\overline{\Lambda_c}\,}}
\def\lc        {\ensuremath{\Lambda_c}}
\def\eelblb      {\ensuremath{e^+ e^- \!\rightarrow \Lambda_b\overline{\Lambda_b}}}
\def\eess        {\ensuremath{e^+ e^- \!\rightarrow \Sigma^0\overline{\Sigma^0}}}
\def\eespsp        {\ensuremath{e^+ e^- \!\rightarrow \Sigma^+\overline{\Sigma^+}}}
\def\eels        {\ensuremath{e^+ e^- \!\rightarrow \Lambda\overline{\Sigma^0}}}
\def\ss        {\ensuremath{\Sigma^0\overline{\Sigma^0}}}
\def\ls        {\ensuremath{\Lambda\overline{\Sigma^0}}}
\def\eenn        {\ensuremath{e^+ e^-\!\rightarrow n\overline{n}\,}}
\def\nn        {\ensuremath{n\overline{n}}}
\def\pp        {\ensuremath{p\overline{p}}}
\def\ll        {\ensuremath{\Lambda\overline{\Lambda}}}
\def\lclc        {\ensuremath{\Lambda_c\overline{\Lambda_c}}}
\def\ee        {\ensuremath{e^+e^-}}
\def\nb        {\ensuremath{{\rm nb}}}
\def\pb        {\ensuremath{{\rm pb}}}
\def\gev       {\ensuremath{{\rm GeV}}}
\def\mev       {\ensuremath{{\rm MeV}}}
\journal{Nuclear Physics A}
\begin{document}
\begin{frontmatter}
\title{Pointlike Baryons?}
%
\author[cf,lnf]{Rinaldo Baldini Ferroli}
\ead{baldini@centrofermi.it}
\author[cf,lnf]{Simone Pacetti\corref{cor1}}
\ead{simone.pacetti@lnf.infn.it}
\author[lnf]{Adriano Zallo}
\ead{adriano.zallo@lnf.infn.it}
\cortext[cor1]{Corresponding author}
\address[cf]{Museo Storico della Fisica e Centro Studi e 
             Ricerche ``E. Fermi'', Rome, Italy}
\address[lnf]{INFN, Laboratori Nazionali di Frascati, Frascati, Italy}
%
%
%
%
%
\begin{abstract}
A peculiar feature, observed in the \bbr\ data on \eebb\ 
cross sections ($\mathcal{B}$ stands for baryon), is the non-vanishing
cross section at threshold for all these processes. This is the expectation  due to the Coulomb 
enhancement factor acting on a charged fermion pair. Remarkably, in the case of \eepp\ it is 
found that Coulomb 
final state interactions largely dominate the cross section at threshold and it turns out a form factor
$|G^p(4 M^2_p)|~\simeq~1$, as a pointlike fermion. 
Also in the case of \eelclc, as recently measured by Belle for the first time,  a
pointlike behavior is suggested for the charmed charged baryon, being the form
factor at threshold $|G^{\lc}(4 M^2_{\lc})| \simeq 1$, even if within a large error.  
In the case of neutral strange baryons the non-vanishing
cross section at threshold is interpreted as a remnant of
quark pair Coulomb interaction before the hadronization, taking into account the 
asymmetry between attractive 
and repulsive Coulomb factors. Besides strange baryon cross sections are successfully compared to U-spin 
invariance relationships. 
\end{abstract}
%
%
%
\end{frontmatter}
\section{$\sigma(\eebb)$ at threshold}
\indent
Unexpected features, observed by \bbr~\cite{pp,llc} in the case of baryon pairs production
and already pointed out \cite{nostre-pub}, are revisited in the following
as well as a further evidence in the charm sector, as recently found by Belle~\cite{belle}. 
They  concern  
cross section measurements at the corresponding threshold energy regions of  
\be
\eepp 
\label{eq:pp}
\en
and
\be
\eell,\; \ss,\; \ls\,.
\label{eq:bb}
\en
\bbr\ has measured the cross section~(\ref{eq:pp}) with unprecedented accuracy and 
the cross sections~(\ref{eq:bb}) for the first time (fig.~\ref{fig:cross-sections}).
Recent results achieved by Belle concerning 
\be
\eelclc \,
\en
are also consistent (fig.~\ref{fig:cs-belle}) with the aforementioned features of \pp.\\
Both, \bbr\ and Belle, have obtained their results
by means of the initial state radiation technique (ISR), 
in particular detecting the photon radiated by the incoming beams. 
There are several advantages in measuring two body processes in this way:
even exactly at threshold the efficiency is quite high, 
a very good invariant mass resolution is achieved ($\sim$1 \mev\ comparable
to symmetric storage rings)
and a full angular acceptance is also obtained, due to the detection of the radiated photon. 
%
%
%
\bfi[h!]
\bc
\epsfig{file=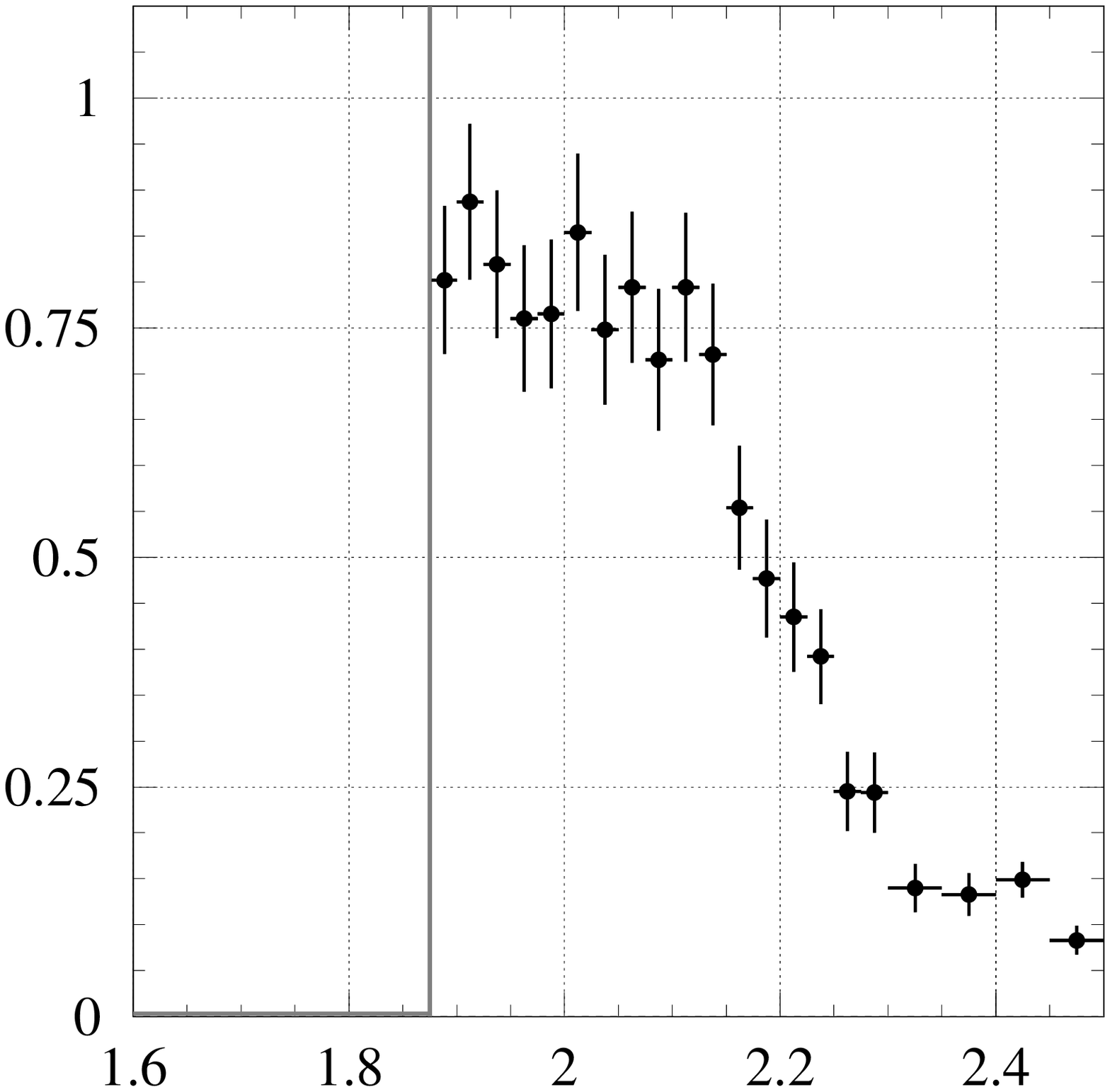,width=60mm}
\put(-52,0){$W_{\pp}$ (GeV)}
\put(-175,84){\rotatebox{90}{$\sigma(\eepp)$ (nb)}}
\put(-115,60){\rotatebox{90}{\boldmath\bf$\pp$ threshold}}
\put(-18,155){\bf a}
\hspace{10mm}
\epsfig{file=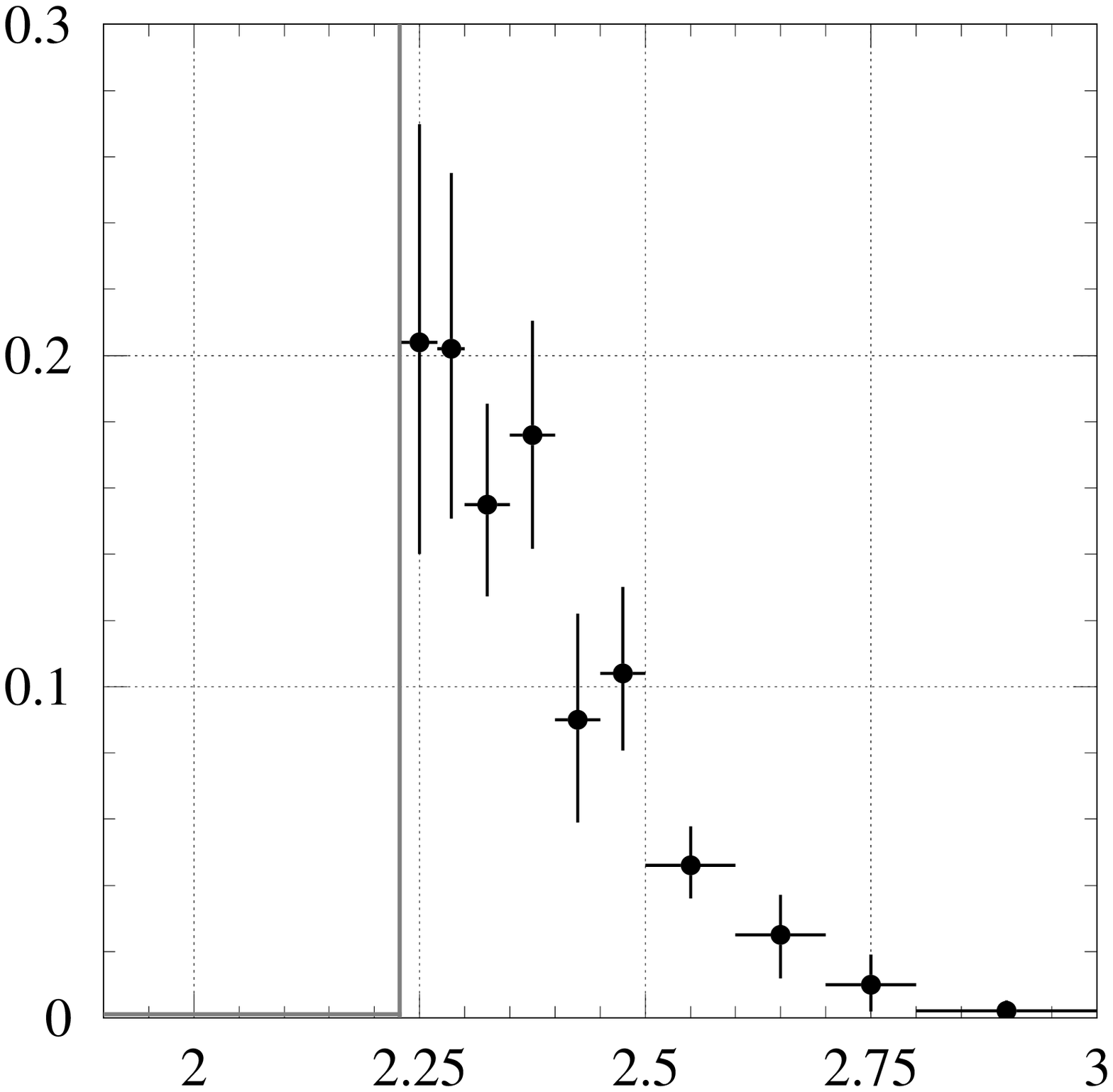,width=60mm}
\put(-55,0){$W_{\ll}$ (GeV)}
\put(-177,82){\rotatebox{90}{$\sigma(\eell)$ (nb)}}
\put(-121,58){\rotatebox{90}{\boldmath\bf$\ll$ threshold}}
\put(-18,155){\bf b}\vspace{2mm}\\
\epsfig{file=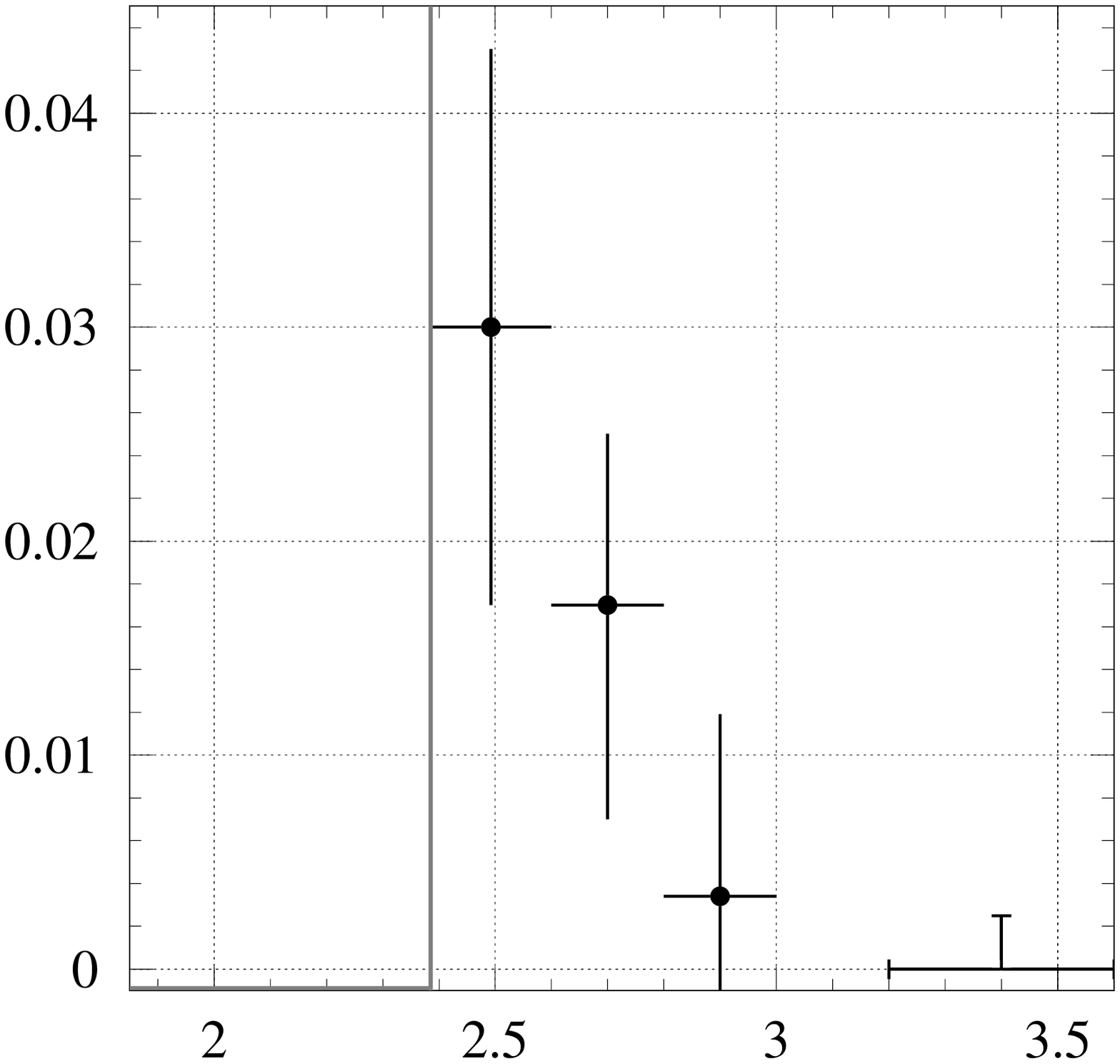,width=60mm}
\put(-63,0){$W_{\ss}$ (GeV)}
\put(-182,72){\rotatebox{90}{$\sigma(\eess)$ (nb)}}
\put(-121,53){\rotatebox{90}{\boldmath\bf$\ss$ threshold}}
\put(-18,155){\bf c}
\hspace{10mm}
\epsfig{file=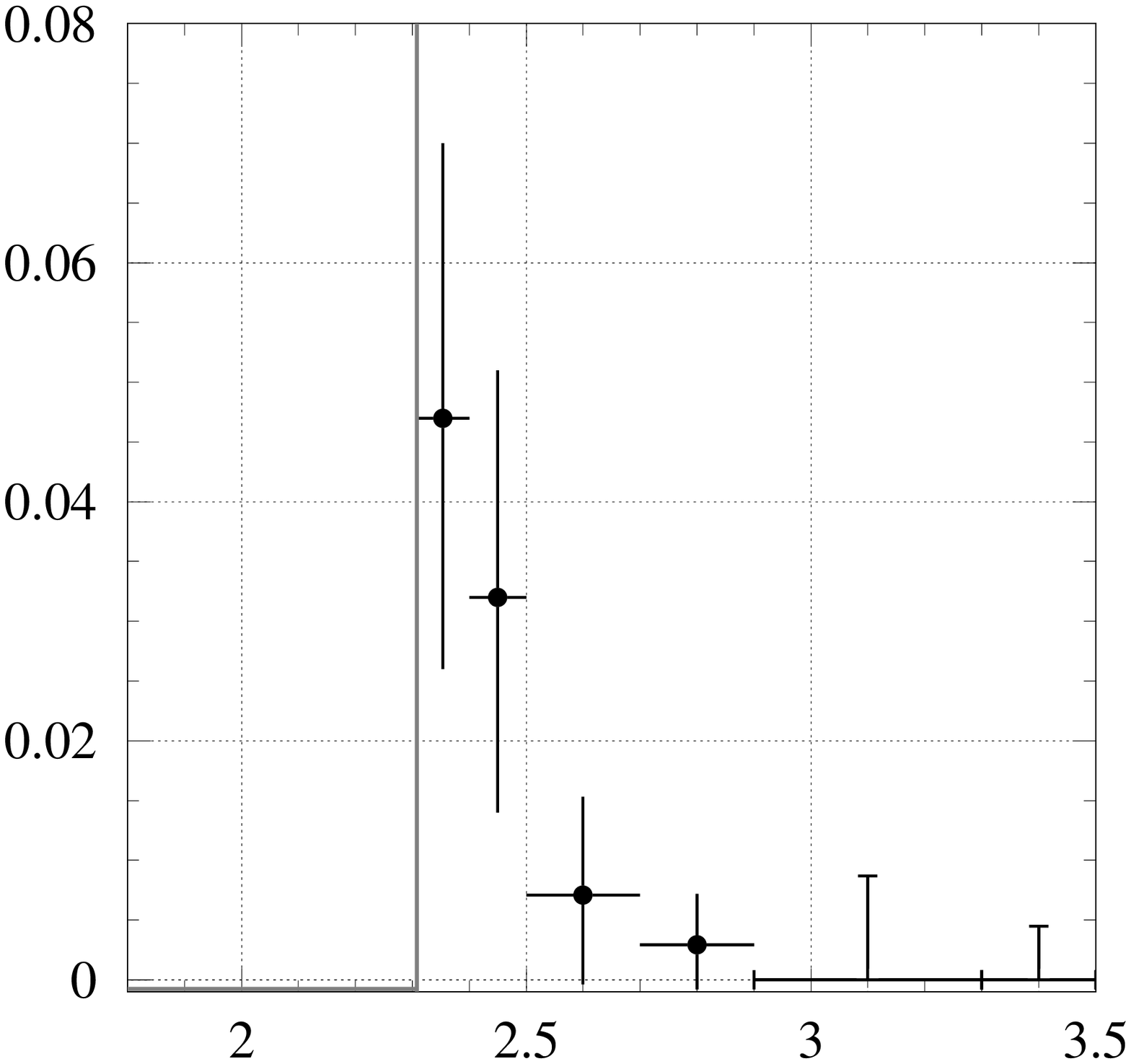,width=60mm}
\put(-59,0){$W_{\ls}$ (GeV)}
\put(-185,76){\rotatebox{90}{$\sigma(\eels)$ (nb)}}
\put(-123,57){\rotatebox{90}{\boldmath\bf$\ls$ threshold}}
\put(-18,155){\bf d}
\ec
\vspace{-5mm}
\caption{\label{fig:cross-sections}%
\eepp\ (a), \eell\ (b), \eess\ (c), and \eels\ (d) total cross sections, as measured 
by \bbr~\cite{pp,llc}. The gray vertical lines indicate the production
thresholds.%
}
\vspace{0mm}
\efi\\
In Born approximation the differential cross section for the process \eebb\ is
\bea
\displaystyle\frac{d\sigma(\ee\!\!\to\bb)}{d\Omega}
\!=\!
\displaystyle\frac{\alpha^2\beta C}{4W_{\bb}^2}\!\left[(1\!+\!\cos^2\theta)
|G_M^\b(W_{\bb}^2)|^2\!+\!\frac{4M_{\b}^2}{W_{\bb}^2}\sin^2\theta|G_E^\b(W_{\bb}^2)|^2\right ],
\label{eq:cross}
\eea
where $W_{\bb}$ is the \bb\ invariant mass, 
$\beta$ is the velocity of the outgoing baryon, $C$ is a Coulomb enhancement 
factor, that will be discussed in more detail in the following, $\theta$ is the 
scattering angle in the center of mass frame and, $G_M^\b$ and $G_E^\b$ are the 
magnetic and electric Sachs form factors (FF).
At threshold it is assumed that, according to the analyticity of the Dirac and Pauli FF's
as well as to the S-wave dominance, there is one FF only: 
$G_E^\b(4 M^2_\b) = G_M^\b(4 M^2_\b) \equiv G^\b(4 M^2_\b)$.
\\
The following peculiar features have been observed, in the case of \eepp~\cite{pp}:
the total cross section $\sigma(\eepp)$ is suddenly 
different from zero at threshold, as it is shown in fig.~\ref{fig:cross-sections}a, 
being $0.85 \pm  0.05$ \nb\ (by the way it is the only endothermic process 
that has shown this peculiarity). 
\\
\bbr\ data on $\sigma(\eepp)$ show a flat behavior, 
within the experimental errors, in an 
interval of about 200 MeV above the threshold and then drop abruptly. The angular distributions show
a dominance of the electric FF $G_E^p$ just above threshold and then a behavior 
driven by the contribution of magnetic FF $G_M^p$.
%
%
%
\bfi[h]\vspace{-0mm}
\bc
\raisebox{4.75mm}{\bm{65mm}
\epsfig{file=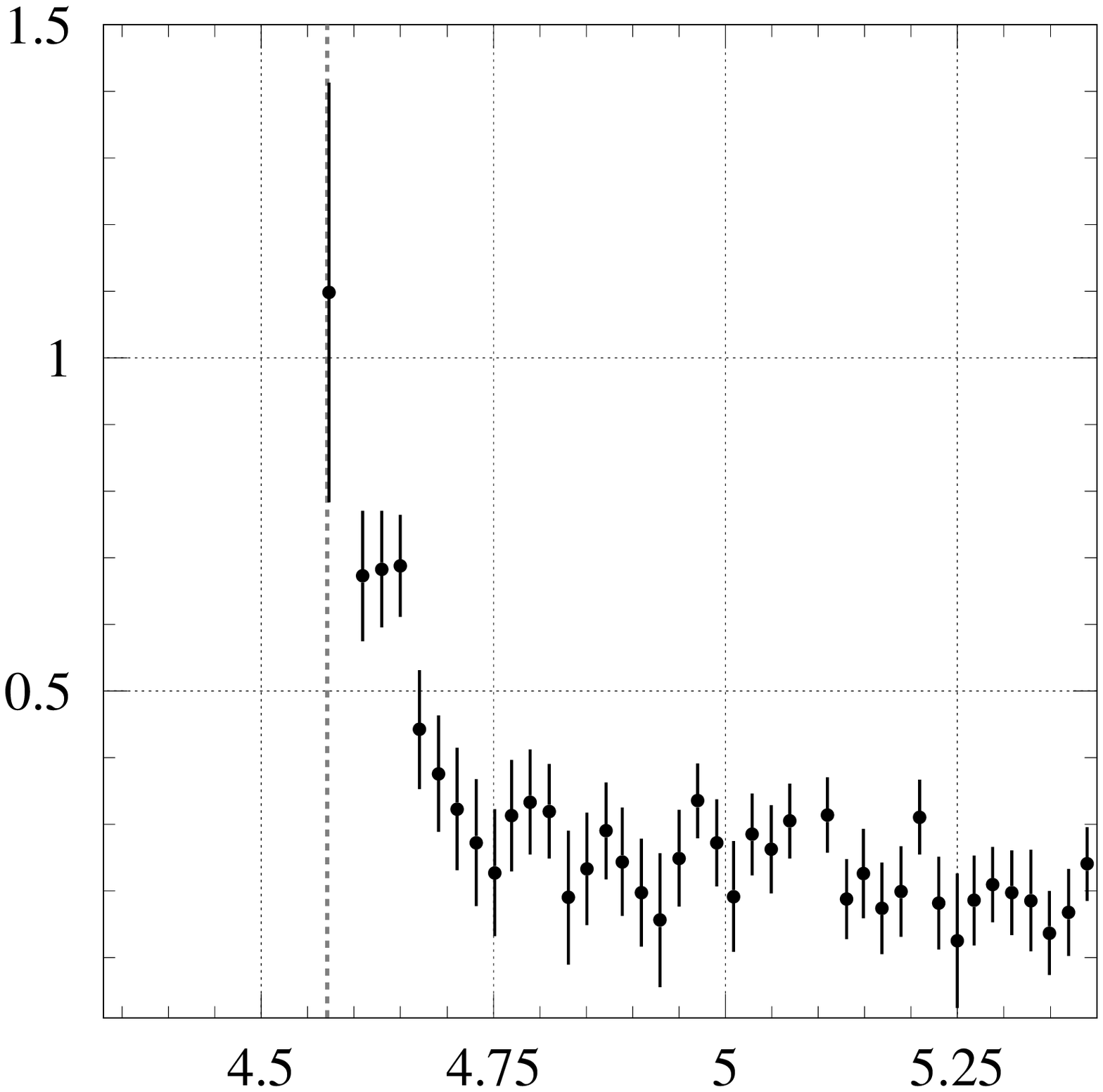,width=65mm}
\put(-63.5,0){$W_{\lclc}$ (\gev)}
\put(-190,159){\rotatebox{90}{$|G^{\lc}|$}}
\put(-140,60){\rotatebox{90}{\boldmath\bf$\lclc$ threshold}}
\vspace{-0mm}
\caption{\label{fig:cs-belle} The \lc\ effective FF, extracted from
the $\sigma(\eelclc)$ cross section measured 
by Belle~\cite{belle}.%
}
\em}\hfill
\bm{65mm}
\epsfig{file=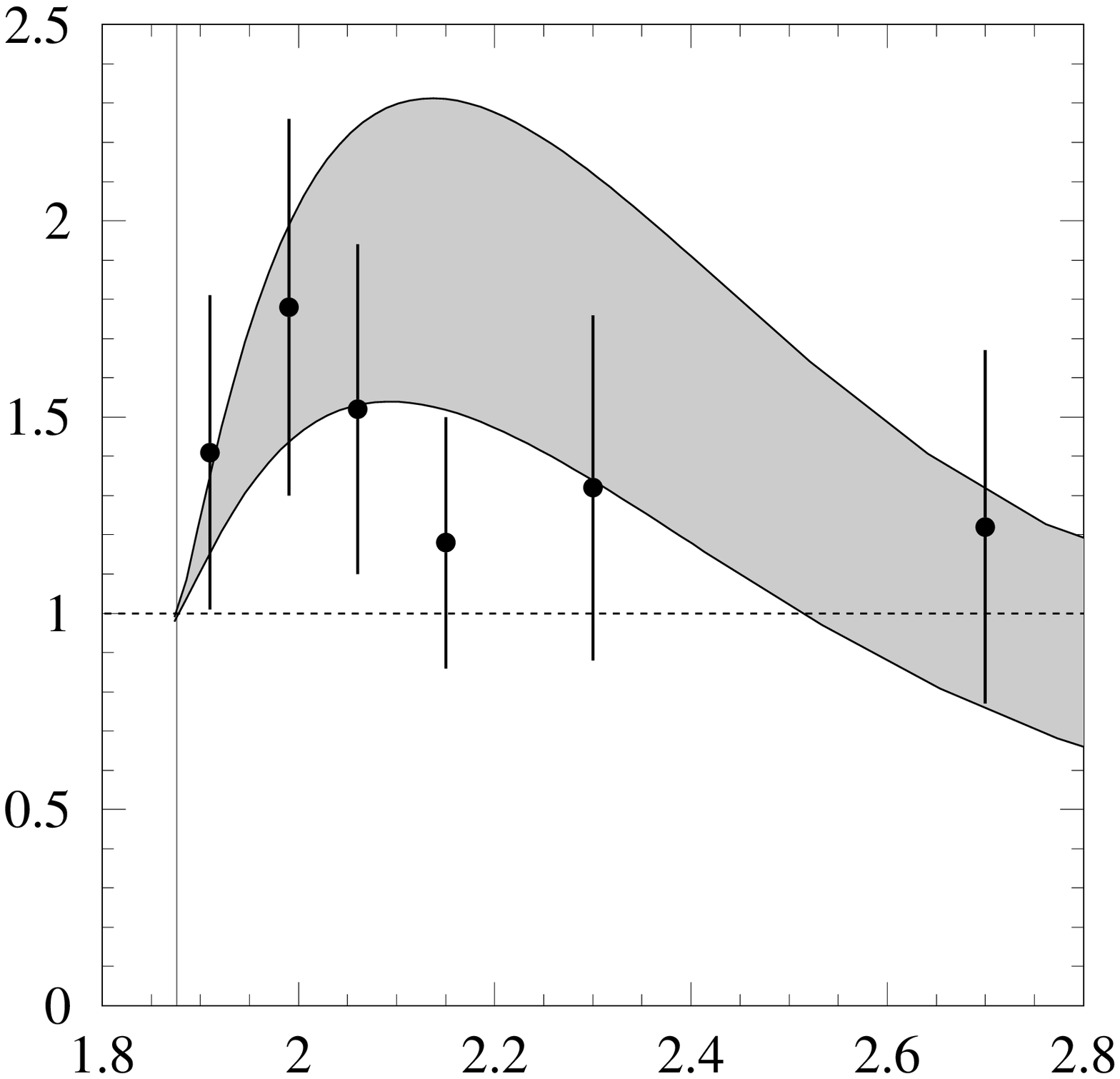,width=65mm}
\put(-53,0){$W_{\pp}$ (\gev)}
\put(-195,139){\rotatebox{90}{$|G_E^p/G_M^p|$}}
\caption{\label{fig:ratio}%
\bbr\ data on the ratio $|G_E^p/G_M^p|$ extracted by studying the angular distribution
of the \eepp\ differential cross section [eq.(\ref{eq:cross})]. The strip is a calculation~\cite{noi} based
on a dispersion relation relating these data and the space-like ratio, as recently achieved
at JLAB and MIT-Bates~\cite{jlab-mitbates}.}
\em
\ec
\efi\\
%
%
Long time ago a final state Coulomb correction to the Born cross section 
has been pointed out in the case of  charged fermion pair production~\cite{coulomb}.
This Coulomb correction is usually introduced as an enhancement factor, $C$ in eq.~(\ref{eq:cross}). 
It corresponds to the squared value of the Coulomb scattering wave function at the origin, assumed as a good 
approximation in the case of a long range interaction added to a short range one, the so called 
Sommerfeld-Schwinger-Sakharov rescattering formula~\cite{coulomb,somm}. This factor has a very weak 
dependence on the fermion pair total spin, hence it is assumed to be the same for $G_E$ and $G_M$ and can be 
factorized. The Coulomb enhancement factor is
\bea
C(W_{\bb}) = \left\{\begin{array}{ll}
1& \mbox{for neutral $\mathcal{B}$}\\
&\\
\displaystyle\frac{\pi\alpha/\beta}{1-e^{-\pi\alpha/\beta}}& \mbox{for charged $\mathcal{B}$}\\
\end{array}\right., \hspace{10mm}
\beta=\sqrt{\displaystyle1-\frac{4M^2_\mathcal{B}}{W_{\bb}^2}}\,.
\label{eq:corr}
\eea
In Ref.~\cite{yogi} a similar formula is obtained, but $1/\beta \rightarrow 1/\beta -1$,
 anyway not affecting the following considerations.
Very near threshold the Coulomb factor is $C(W_{\bb}^2\to 4M_\mathcal{B}^2)\simeq\pi \alpha /\beta$,  
so that the phase space factor $\beta$ is cancelled and the cross 
section is expected to be finite and not vanishing even exactly at threshold. 
However, as soon as the fermion relative velocity 
is no more vanishing, actually few MeV above the threshold, it is $C \simeq 1$ and Coulomb effects can 
be neglected.
\\
In the case of \eepp\ the expected Coulomb-corrected cross section at threshold is
\bea
\sigma(\eepp)(4M_p^2) = \frac{\pi^2\alpha^3}{2M_p^2} \cdot |G^p(4M_{p}^2)|^2  
\simeq 0.85\cdot |G^p(4 M^2_p)|^2 \;\nb\,, \no
\eea
in striking similarity with the measured one. 
Therefore the Coulomb interaction dominates the energy region 
near threshold and it is found
\be
\fbox{\boldmath$|G^p(4 M^2_p)| = 0.97\pm 0.04$(stat)$\pm0.03$(syst)} \,.\no
\en
That is, a \pp\ pair produced at threshold behaves like a pointlike fermion pair. 
In the following this feature is suggested
to be a general one for baryons. It looks as if the FF at threshold, interpreted as
$\mathcal{B}$ and $\overline{\mathcal{B}}$ wave function static overlap, 
coincides with the baryon wave function normalization, taking into account S-wave 
is peculiar of fermion pairs at threshold. 
In the case of meson pair production, the 
total angular momentum conservation requires a P-wave, that vanishes
at the origin as well as the quoted Coulomb enhancement factor and the 
cross section has a $\beta^3$ behavior near threshold. Tiny Coulomb effects in the 
case of meson pairs have been extensively pursued~\cite{Voloshin}. 
%
%
%
\bfi[ht]
\begin{flushright}
\epsfig{file=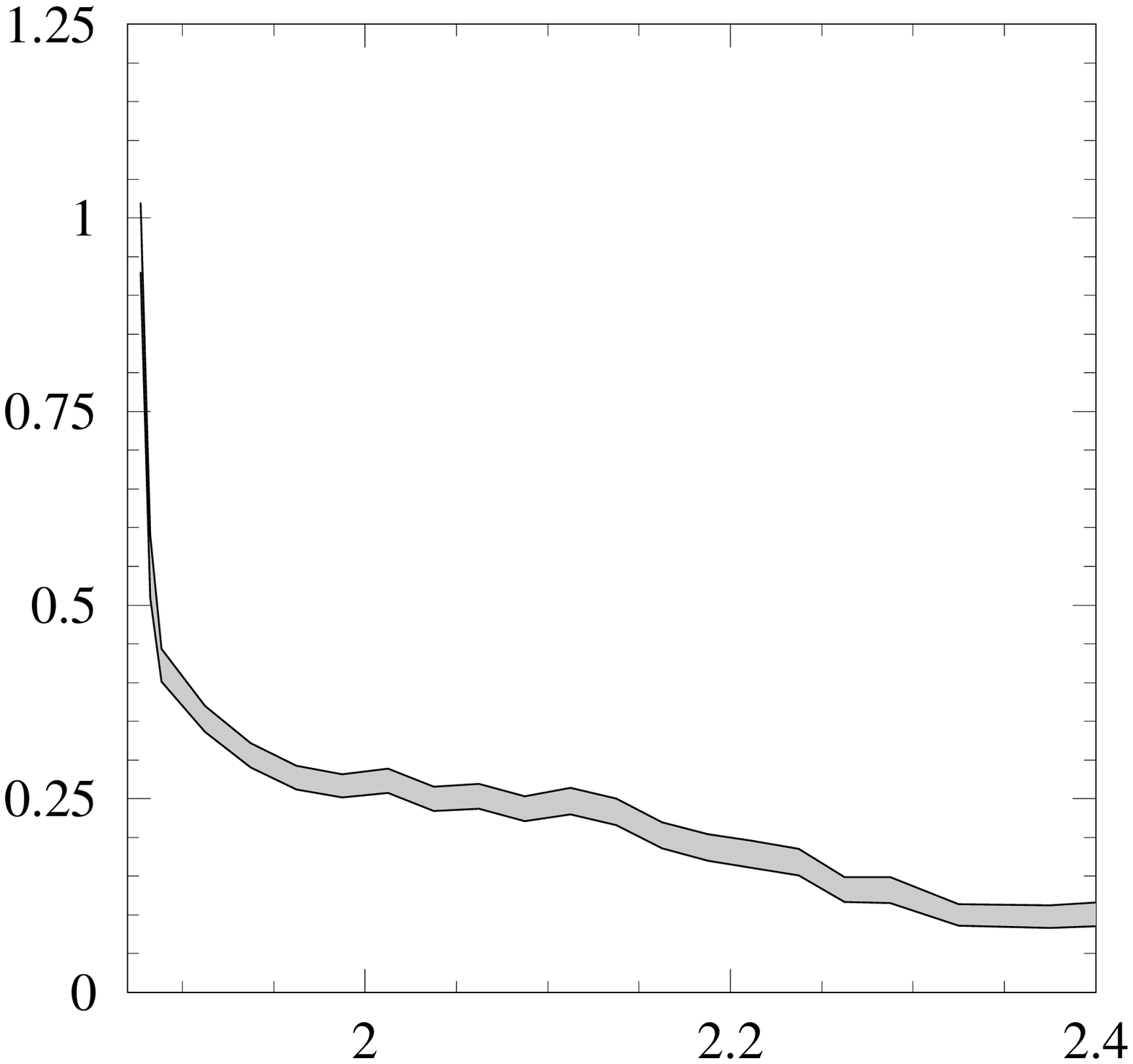,width=63mm}
\put(-53,0){$W_{\pp}$ (GeV)}
\put(-188,131){\rotatebox{90}{$|B_S^p(W_{\pp}^2)|$}}
\put(-20,161){\bf a}
\hspace{4mm}
\epsfig{file=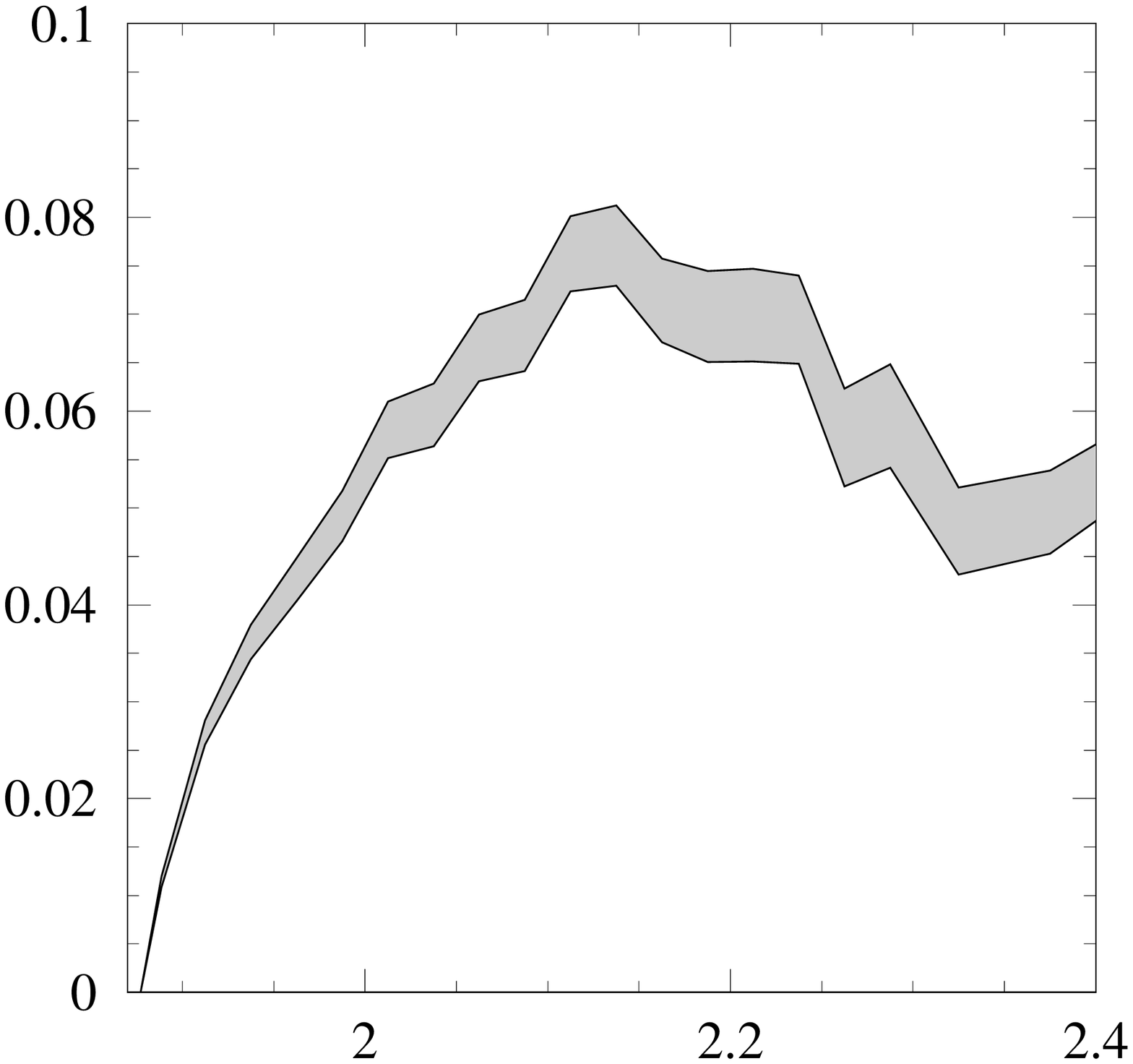,width=63mm}
\put(-53,0){$W_{\pp}$ (GeV)}
\put(-188,129.5){\rotatebox{90}{$|B_D^p(W_{\pp}^2)|$}}
\put(-20,161){\bf b}
\end{flushright}
\vspace{-5mm}
\caption{\label{fig:waves}%
S-wave (a) and D-wave (b) FF's as obtained in an updated version of Ref.~\cite{noi}, from a 
dispersive analysis based
on the \bbr\ data on the total \eepp\ cross section and the time-like ratio $|G_E^p/G_M^p|$.
The point at threshold, included in the fit, is 
exactly zero for the D-wave (b), while it is $\simeq 1$ for the S-wave (a).
}
\efi
Angular momentum and parity conservation allow, in addition to 
the S-wave, also a D-wave contribution.
The angular distribution, averaged in a 100 MeV interval above the threshold, has a behavior 
like $\sin^2\theta$, i.e. dominated by the electric FF, and then a behavior like 
$(1+\cos^2\theta)$, i.e. driven by the magnetic FF contribution [see eq.~(\ref{eq:cross}) and fig.~\ref{fig:ratio}].
In particular the different behavior at threshold and the dominance of the electric FF are consistent with
a sudden and large D-wave contribution.
 In Ref.~\cite{noi} the relative phase and therefore the S- and D-wave complex FF's, 
$B_S^p$ and $B_D^p$, have been extracted, by means of a dispersion relation, applied
to the space-like ratio $G_E^p/G_M^p$ and to the \bbr\ time-like $|G_E^p/G_M^p|$
(fig.~\ref{fig:ratio}). 
The S-wave has a sharp drop above threshold, as shown in fig.~\ref{fig:waves}, 
consistent with a Coulomb dominance. The D-wave vanishes at the origin,
it is not affected by the Coulomb enhancement factor and the corresponding
D-wave only cross section should behave as $\beta^5$ at threshold.
S-wave and D-wave  have opposite trends, producing the observed plateau in the total cross section.
Actually, since Coulomb corrections affect the S wave rather than the D wave, the
cross section should be written in terms of $\sqrt(C) B_S$ and $B_D$.
The outcome is under investigation~\cite{inprogress}.
\\
The aforementioned arguments can be tested in the case of
\eespsp\ and it 
should be at threshold, in agreement with the U-spin expectation too:
$\sigma(\eespsp) \simeq \sigma(\eepp) \cdot (M_{p}/M_{\Sigma^+})^2 \simeq 0.53\, \nb.$
This measurement has not yet been done, 
but it is within the \bbr\ and Belle capabilities by means of ISR.
\vspace{3mm}
\\
Recently Belle has measured for the first time the \eelclc\ cross section~\cite{belle}.
A resonant behavior, just above threshold, has been pointed out. The anomalously high
cross section close to the \pp\ one, prefigures a similar FF.
Still the cross section is not zero at threshold.
In fig.~\ref{fig:cs-belle} the $G^{\lc}$ effective FF is reported, obtained under the 
hypothesis that $|G_E^{\lc}| = |G_M^{\lc}|$, as it is expected at threshold. 
That is the best one can do lacking
angular distributions.
The FF at threshold is achieved by taking the mean value of the first and the second
point of the \lclc\ cross section given in Ref.~\cite{belle}. This procedure or similar
recipes to cure the threshold behavior must be used because, due to the finite 
energy-resolution, the events at threshold are spread in an energy bin that extends itself 
below $W_{\lclc}=2M_{\lc}$.
A systematic uncertainty of 30\% is estimated for this procedure, to be added in quadrature to the
other systematic errors.
Taking into account the Coulomb correction, the statistical errors and the overall 
systematic quoted errors 
added in quadrature,
mostly due to the uncertainties in the \lc\ branching ratios,
it is found:
\be
|G^{\lc}(4 M^2_{\lc})| = 1.1 \pm 0.3 (\rm stat) \pm 0.4 (syst)\no\,.
\en
That is close, in a tantalizing way, to the aforementioned suggested feature: 
 baryon pairs, produced at threshold, behave as pointlike fermions.
The hypothesis it is a general feature for baryons is strengthened
by the gap between $4 M^2_{\lc}$ and $4 M^2_p$.
Extrapolating to \eelblb\ it should be at threshold $\sigma(\eelblb) \simeq 23\, \pb$, which
means one of the most important hadronic channels at those energies.
\section{The puzzle of neutral baryon form factors}
The peculiar features of \eepp\ have been observed by \bbr~\cite{llc} also in the case of \eell, \ss, \ls\ 
(fig.~\ref{fig:cross-sections}b, c, d), even if within bigger experimental errors. In particular 
the cross section  $\sigma(\eell)$ is different from zero at threshold, being $0.20~\pm~0.05$~nb.
The cross sections $\sigma(\eess)$ and $\sigma(\eels)$ have been measured by  
\bbr\  for the first time. At threshold,
assuming a smooth extrapolation from the first data point, it is 
$\sigma(\eess)=0.03\pm 0.01$ nb and $\sigma(\eels)=0.047\pm 0.023$ nb. 
In all these cases final state Coulomb effects should 
not be taken into account and a finite cross section at threshold is not expected. 
Nevertheless, in the case of the neutral $\Lambda$ baryon, not only the 
cross section data, but also the ratio $|G_E^\Lambda/G_M^\Lambda|$ have a trend similar to
$|G_E^p/G_M^p|$.
\\
One might investigate what is expected in the debatable hypothesis a two steps process occurs:
at first unbound quark pairs are produced coherently 
and then the hadronization takes place. Hence the finite cross section at threshold
should be a remnant of the initial step. 
Valence quarks only are considered in the following. 
For each quark pair there is a Coulomb attractive 
amplitude times the quark electric charge and each amplitude 
has a phase taking into account the displacement of the quark pair inside the baryon. 
In addition to the quark pair Coulomb interaction there are contributions from quarks
belonging to different pairs. However there are several suppression factors for them:
relative phase, velocity spread and moreover most of them, coming from quarks having 
charges of the same sign, are repulsive ones. There is no symmetry between  repulsive and attractive 
Coulomb interactions and this asymmetry might explain why there is 
a non-vanishing cross section at threshold  even for neutral baryon pairs. 
In fact in the case of repulsive Coulomb interaction the Sommerfeld 
formula is (charges $Q_{q}$ and $Q_{\ov{q}'}$ have the same sign):
\bea
C(W_{\bb}) =\frac{-\pi\alpha |Q_{q}Q_{\ov{q}'}|/\beta}{1-\exp(+\pi\alpha|Q_qQ_{\ov{q}'}|/\beta)}
\mathop{\longrightarrow}_{W_{\bb}^2\to4M_{\b}^2}0 \no\,,
\eea
i.e. $C=0$ at threshold.
Considering only Coulomb enhancement factors due
to quark pairs, the same cross section is expected in the \pp\ case:
\bea
\sigma(\eepp)(4M_p^2) = \frac{\pi^2\alpha^3}{2M_p^2}(2Q_u^2 + Q_d^2) 
\simeq 0.85  \,\nb\,, \no
\eea
while, for instance in the $\Lambda$ case, it is:
\bea
\sigma(\eell)(4M_\Lambda^2)= 
\frac{\pi^2\alpha^3}{2M_\Lambda^2}(Q_u^2 + Q_d^2 + Q_s^2)  
\simeq  0.4 \; \nb\,. \no
\eea  
Hence in the case of $\eell$ the 
expectation range is $(0- 0.4)~\nb$ to be 
compared to the experimental value at threshold: $\sigma(\eell)= 0.20 \pm 0.05$ nb.
\\
According to the U-spin invariance~\cite{park}, assuming a negligible 
electromagnetic transition between U-spin triplet and singlet,
strange baryon magnetic FF's are related, being:
\bea
G_{\Sigma^0} - G_\Lambda + \frac{2}{\sqrt{3}}G_{\ls} = 0 \,.
\label{eq:uspin-ff}
\eea
\\
In terms of adimensional quantities there is a good agreement with
the \bbr\ results, assuming real same-sign FF's at threshold:
\bea
M_{\Sigma^0}\sqrt{\sigma_{\ss}}-M_\Lambda\sqrt{\sigma_{\ll}}+
\frac{2}{\sqrt{3}}\ov{M_{\ls}}\sqrt{\sigma_{\ls}}=(-0.1\pm2.0)\times 10^{-4}. \no
\eea
The \ss\ cross section at threshold, due to the aforementioned relationship 
with \eell\ and \eels, is predicted to be: $\sigma_{\ss} = 0.03 \pm 0.03\, \nb $.
\\
Vector mesons poles in the unphysical region below threshold ($Q^2 < 4 M_\mathcal{B}^2$) 
largely account for the
cross sections above threshold. An extrapolation of the strange baryon magnetic moments (i.e. $Q^2 = 0$)
is not expected  a priori. However, if there is no sign change in the space-like
region, analyticity requires time-like asymptotic FF's should have at least the same sign. 
Therefore the \eell\ and the \eels\ amplitudes are supposed to be both negative
in agreement with the magnetic moments. The \eess\ turns out to be also
negative, in fair disagreement with this statement.
\bfi[ht]\vspace{-0mm}
\bc
\epsfig{file=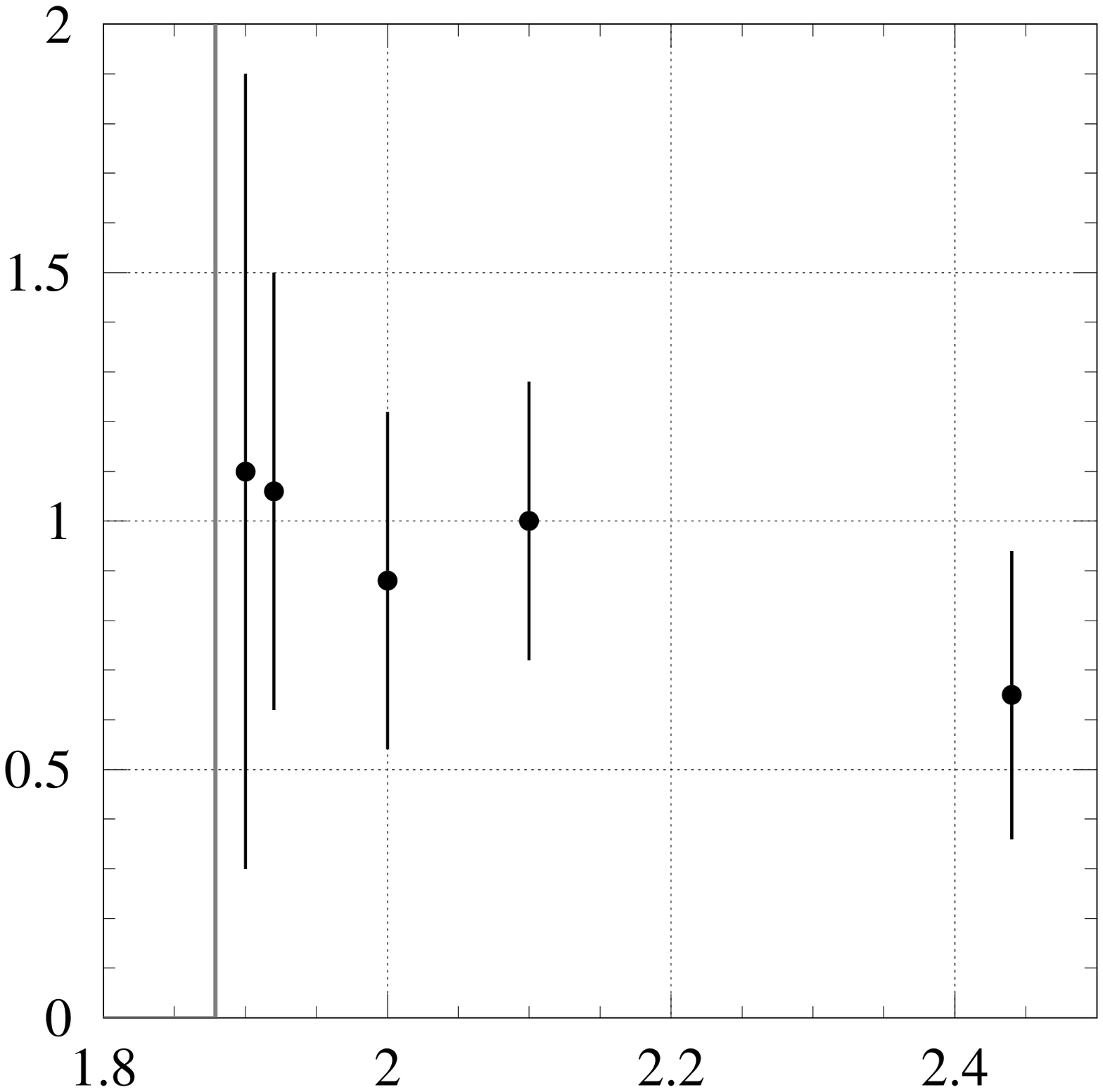,width=75mm}
\put(-55,3){$W_{\nn}$ (GeV)}
\put(-218,127){\rotatebox{90}{$\sigma(\eenn)(\nb)$}}
\vspace{-2mm}
\caption{\label{fig:s0s0nn}%
The \eenn\ total cross section as measured by
the FENICE Collaboration~\cite{adone}.}
\ec
\efi\vspace{-0mm}
\\
An important process to understand the neutral baryon puzzle is
\eenn. 
The cross section $\sigma(\eenn)$ has been measured only once, long time ago by  
FENICE  at the $e^+ e^-$ storage ring ADONE~\cite{adone} and it was found  
$\sigma(\eenn)\!\simeq$~1~nb, as shown in fig.~\ref{fig:s0s0nn}.
According to the above mentioned assumption on U-spin invariance it should be
$G_n=\frac{3}{2}G_\Lambda-\frac{1}{2}G_{\Sigma^0}$,
hence
\bea
\sigma(\eenn)=\frac{1}{4}\left(3\sqrt{\sigma_{\ll}}M_\Lambda-
\sqrt{\sigma_{\ss}}M_{\Sigma}\right)^2\frac{1}{M_n^2}=
0.5\pm0.2\,\nb\,.
\label{eq:nn}
\eea
Unfortunately it is very unlike that \bbr\ or Belle will ever be able to measure this
process by means of the ISR, but results on this process 
are expected by BESIII~\cite{BESIII} and VEPP2000~\cite{VEPP2000}. 
\\
It has to be reminded that various theoretical models and phenomenological descriptions have made
predictions on baryon time-like FF's~\cite{models}. 
%
\section{Conclusions}
\label{sec:conclu}
All the \eebb\ cross sections, as measured by  \bbr, do not vanish at threshold.
In the case of \eepp\ this behavior is explained by the \pp\, Coulomb enhancement factor 
and it comes out that $|G^p(4 M^2_p)| \simeq 1$, that is proton pairs behave as pointlike
fermions. This cross section is remarkably flat near threshold: it turns out that S- and 
D-wave have opposite trends, producing this peculiar behavior and the S-wave contribution 
has a steep drop above threshold,
consistent with Coulomb dominance.
A pointlike behavior is suggested also in the case of the \lc\ FF at threshold, as 
recently achieved by Belle.
Neutral strange baryons show a non-vanishing cross section at threshold too, which might be interpreted
as a remnant of quark pair Coulomb interaction before hadronization.
A consistent framework
of strange baryon FF's is obtained
requiring the suppression of electromagnetic transitions between U-spin singlet and
triplet.
Neutron and $\Sigma^+$ FF's are demanded to check this picture of baryon FF's.
\section*{Acknowledgments}
We owe special thanks to Antonino Zichichi, who pioneered and still leads the field of
baryon FF, to Stan Brodsky and Yogi Srivastava, who suggested to look more carefully to
Coulomb-like enhancement factors, and warmly acknowledge \bbr\ physicists from the Budker Institute, 
who achieved the aforementioned \bbr\ cross sections. 
%
%
%
%

\end{document}